# Research on Ultrasonic Imaging Systems Based on Scattering Field Theory

Kai Yabumoto[1]*, Takayoshi Yumii[2] and Kenjiro Kimura[1,2]

**AFFILIATIONS**

[1]*Kobe University, 1-1, Rokkodai-cho, Nada-ku, Kobe-shi, Hyogo, 657-8501, Japan*

[2]*Integral Geometry Science, 1-5-6, Minatojima-minamimachi, Chuo-ku, Kobe-shi, Hyogo,*

*650-0047, Japan*

*The author to whom correspondence may be addressed: kai@ig-instrum.co.jp

**Keyword:** Orthogonal linear arrays, air-coupled ultrasonic sensors, scattering field theory, analytical solution

In an ultrasonic array system, increasing the aperture size to achieve a high resolution requires more transmit and receive channels, thus making it essential to have an analysis technique that can reconstruct the shape and physical properties of scatterers in a finite time based on measurement data obtained from numerous ultrasonic elements. Herein, we developed an ultrasound imaging technique using a two-dimensional ultrasonic array system comprising $N^2$ transmitter/receiver pairs and a technique for controlling it with a *2N*-channel transmitter/receiver circuit. In addition, we derived a partial differential equation that describes wave propagation in the scattering field where the transmitter and receiver arrays are orthogonally arranged, based on scattering field theory. By analytically solving this equation, we derived an imaging function that incorporates the measurement data as boundary conditions. Moreover, we visualized the spatial distribution of reflection intensity corresponding to the target shape by measuring reflected waves from the target using this measurement system and performing reconstruction calculations based on scattering field theory. We demonstrated the feasibility of ultrasonic imaging by combining ultrasonic measurements using a smaller number of channels than the number of ultrasonic elements and scattering field theory and using the measurement results as boundary conditions.

**Introduction**

In ultrasonic array devices, when the aperture size is increased to obtain a high resolution, the number of input/output channels increases, thus necessitating an analysis technique that enables the reconstruction of the shape and physical properties of scatterers in a finite time from measurement data obtained from numerous ultrasonic elements. Herein, we constructed an ultrasound imaging system that combines an imaging function derived from the scattering field theory with an ultrasound measurement device using a bistatic method (hereinafter referred to as orthogonal linear array) in which one-dimensional linear arrays are orthogonally arranged. In the orthogonal linear array method, the total number of circuit connections from the two-dimensional array transducer to the switching circuit or amplification circuit can be configured as $2N$ channels. This overcomes the issue of the number of channels[1] required for the circuit connections ($N^2$) for conventional two-dimensional arrays and has the advantage of making it easier to expand the aperture size of the two-dimensional array. In addition, the energy of the transmitted and received signals increases as the aperture size increases, thereby deepening the penetration depth and improving the signal-to-noise ratio (SNR)[2, 3]. However, beamforming used in conventional orthogonal linear array imaging systems to form a focus[4, 5] requires the phase delay calculations for every voxel in the imaged region[2, 6-10]. By contrast, we used scattering field theory[11-13] to derive the differential equations of the measurement system using the orthogonal linear array method and derived the imaging function. By employing this analytical solution, we developed a measurement system that can reconstruct tomographic images in the measurement area from the measurement results of the orthogonal linear array method.

**Methods**

In this section, we describe the details of the ultrasonic imaging system developed in this study. Figure 1 shows a schematic of the measurement system comprising ultrasonic transmitters, receivers, and a target. One-dimensional (1D) transmitter arrays and receiver arrays are orthogonally arranged. Ultrasonic waves emitted from the active transmitter array propagate in a fan shape from the 1D array[14]. The ultrasonic waves reflected from the target are observed through the active receiver array. This measurement event is performed for all combinations () of the transmitter arrays ($N$ rows) and the receiver arrays ($N$ columns). This measurement principle enables the measurement of data points using circuits. From the measurement results, the target structure is visualized using reconstruction theory.

Next, the imaging function is derived. Figure 2 shows the mathematical concept of the ultrasonic measurement device that uses the orthogonal linear array method developed in this study. $\phi$ is a function of wave phase and amplitude; $x_1$ is the row coordinate of the transmitter array; $x_2$ is the coordinate of a point on the receiver array; $y_1$ is the column coordinate of the receiver array; $y_2$ is the coordinate of a point on the transmitter array; $z$ is the depth coordinate; $k$ is the ultrasonic wave number; and $\alpha, \beta$ is the phase difference between adjacent points. Based on the scattering field theory in 11-13, the partial differential equation representing the measurement system in Figure 2 is expressed by Equation (1). Here, to avoid complexity, the differential symbols are abbreviated following the references.

$$\left\{\Delta_5^2 + 4k^2\partial_z^2 - 4\left(\partial_{x_1}^2 + \partial_v^2\right)\left(\partial_u^2 + \partial_{y_1}^2\right)\right\}\phi = 0 \tag{1}$$

The signal obtained in the measurement is owing to backscattering. Therefore, the general solution of Equation (1) is Equation (2).

$$\phi(x_1, y_1, x_2, y_2, z, k) = \frac{1}{(2\pi)^4} \int_{-\infty}^{\infty}\int_{-\infty}^{\infty}\int_{-\infty}^{\infty}\int_{-\infty}^{\infty} e^{-i(k_{x_1}x_1 + k_{y_1}y_1)} e^{-i(k_{x_2}x_2 + k_{y_2}y_2)} a(k_{x_1}, k_{y_1}, k_{x_2}, k_{y_2}, k)$$
$$\cdot e^{iz\left\{\sqrt{k^2 - \left\{k_{x_1}^2 + (k_{y_2} + \beta k)^2\right\}} \pm \sqrt{k^2 - \left\{k_{y_1}^2 + (k_{x_2} + \alpha k)^2\right\}}\right\}} dk_{x_1} dk_{y_1} dk_{x_2} dk_{y_2} \tag{2}$$

Based on the measurement principle of this study, Equation (3) is obtained by calculating the sum of $x_2$ and $y_2$ and rearranging it.

$$\Phi(x_1, y_1, z, \alpha, \beta, k) = \frac{1}{(2\pi)^2} \int_{-\infty}^{\infty}\int_{-\infty}^{\infty}\int_{-\infty}^{\infty}\int_{-\infty}^{\infty} e^{-i(k_{x_1}x_1 + k_{y_1}y_1)} \delta(k_{x_2}\Delta x)\delta(k_{y_2}\Delta y) a_{\alpha,\beta}(k_{x_1}, k_{y_1}, k_{x_2}, k_{y_2}, k)$$
$$\cdot e^{iz\left\{\sqrt{k^2 - \left\{k_{x_1}^2 + (k_{y_2} + \beta k)^2\right\}} \pm \sqrt{k^2 - \left\{k_{y_1}^2 + (k_{x_2} + \alpha k)^2\right\}}\right\}} dk_{x_1} dk_{y_1} dk_{x_2} dk_{y_2}$$
$$= \frac{1}{(2\pi)^2 \Delta x \Delta y} \int_{-\infty}^{\infty}\int_{-\infty}^{\infty} e^{-i(k_{x_1}x_1 + k_{y_1}y_1)} b(k_{x_1}, k_{y_1}, \alpha, \beta, k)$$
$$\cdot e^{iz\left\{\sqrt{k^2 - \left\{k_{x_1}^2 + \beta^2 k^2\right\}} \pm \sqrt{k^2 - \left\{k_{y_1}^2 + \alpha^2 k^2\right\}}\right\}} dk_{x_1} dk_{y_1} \tag{3}$$

Furthermore, when the two-dimensional Fourier transform data of the measurement data are defined as $\tilde{Q}(k_{x_1}, k_{y_1}, \alpha, \beta, k)$, Equation (3) can be rearranged as Equation (4).

$$\Phi(x_1, y_1, z, \alpha, \beta, k) = \frac{1}{(2\pi)^2} \int_{-\infty}^{\infty}\int_{-\infty}^{\infty} e^{-i(k_{x_1}x_1 + k_{y_1}y_1)} \tilde{Q}(k_{x_1}, k_{y_1}, \alpha, \beta, k)$$
$$\cdot e^{iz\left\{\sqrt{k^2 - \left\{k_{x_1}^2 + \beta^2 k^2\right\}} \pm \sqrt{k^2 - \left\{k_{y_1}^2 + \alpha^2 k^2\right\}}\right\}} dk_{x_1} dk_{y_1} \tag{4}$$

From the measurement principle in this study, $\alpha = 0, \beta = 0$ is substituted. Moreover, considering the bandwidth of the ultrasonic signal, the analytical expression $\rho(x_1, y_1, z)$ for the 3D reflection intensity distribution can be obtained.

$$\rho(x_1, y_1, z) = \frac{1}{(2\pi)^2} \int_0^\infty \int_{-\infty}^\infty \int_{-\infty}^\infty e^{-i(k_{x_1} x_1 + k_{y_1} y_1)} \tilde{Q}(k_{x_1}, k_{y_1}, k) \cdot e^{i\left(\sqrt{k^2 - k_{x_1}^2} \pm \sqrt{k^2 - k_{y_1}^2}\right)z} dk_{x_1} dk_{y_1} dk \qquad (5)$$

To experimentally verify the above theory, the following measurement system was constructed. Figure 3 illustrates a schematic of the orthogonal linear array type ultrasonic imaging system developed in this study. The system is configured to perform reconstruction imaging based on the imaging function of Equation (5) derived from the scattering field theory. The system comprises a function generator, an orthogonal linear array transducer, a switching circuit, a signal amplifier circuit, a microcontroller, a detection circuit, an A/D converter, and a computer.

The orthogonal linear array transducer uses an aerial ultrasonic sensor (MA40S4S/R, center frequency 40 kHz)[15], which has characteristics useful for principle verification. It is an orthogonal linear array system in which the transmitter array (row direction) and receiver array (column direction) are each one-dimensionally arranged. The transmitter array is driven by a 500-mV$_{pp}$ square wave from a Tektronix AFG1022 function generator via an INA110KP-based amplifier circuit (20 dB gain), with all elements emitting ultrasonic signal in-phase. The receiver array detects the reflected wave. The signals are detected using an INA110KP-based amplifier circuit (40 dB gain).

The switching circuit uses an ADG732 multiplexer, and the transmitting/receiving channels that can be selected for each row and column are controlled by a 5-bit digital signal. This structure allows to perform the measurements of $N^2$ the combinations of transmitters and receivers using circuit connections for a total of channels. The measured data can be substituted into the imaging function in Equation (5) derived by the scattering field theory to visualize the reflection intensity distribution at any depth $z$ mm. Therefore, the amount of calculation required for image reconstruction is $N_{rec} = (N^2) \log_2(N^2)$. The combination of circuit connections of $2N$ channels and $N^2$ data points allows for the scalability of the visualization area and computational efficiency.

**Result and discussion**

An experiment was conducted to visualize multiple targets resembling the letters of the alphabet. In the experiment, a letter-like pattern made of hard foam material placed 340 mm directly above the array was used as the target. The letter-like pattern comprised four letters "K," "O," "B," and "E" with a minimum width of 30 mm (Figures 4 and 5 (a)). In the experiment, equivalent measurements were achieved by scanning the 16 × 16 element array four times at different positions as an alternative to the 32 × 32 element array. Figure 5 (a) shows the size of the letters cut out of the foam material; (b) shows an optical photograph of the letters taken from the front; (c) and (d) show the center frequency (40 kHz) components of the orthogonal biphasic data of the reflected signal obtained experimentally in the setup shown in Figure 4; and (e) shows the absolute value. Figure 5 (d) shows an image obtained by calculating the reflected intensity distribution at z = 340 mm using the imaging function in Equation (5). Figure 5 (c)–(f) shows unclear images with attenuated spatial frequencies, whereas the reconstructed images clearly visualize the letters "K," "O," "B," and "E."

**Conclusion**

Herein, we developed an ultrasonic imaging device using an orthogonal linear array method, in which a two-dimensional ultrasonic array system comprising $N^2$ ultrasonic sensors is efficiently controlled by a 2$N$-channel transmit/receive circuit. Furthermore, by analytically solving the partial differential equation describing the measurement system of this device based on the scattered field theory, we derived an imaging function that does not require beamforming. A clear reflection intensity distribution image was obtained using the proposed method in an experiment using an object cut out of foam material. The results indicate that the issues of circuit complexity and computational load owing to the increasing number of channels to obtain higher resolution images can be solved.

Future study will include the development of an imaging system employing piezoelectric elements suitable for measurement in environments such as underwater and on metal. We also plan to verify the effectiveness of the system for deploying it in applied fields such as biomedical imaging diagnosis and nondestructive testing.

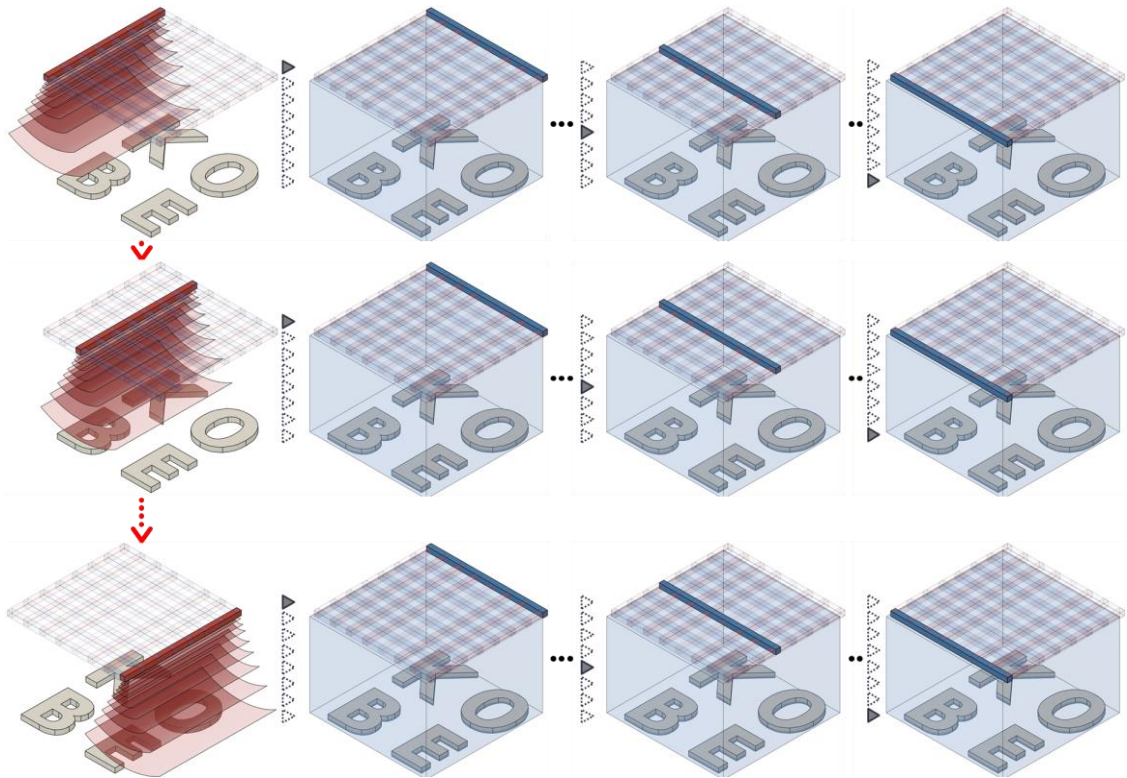

Figure 1 Illustration of the measurement principle of the ultrasonic imaging system. One-dimensional (1D) transmitter arrays and receiver arrays are orthogonally arranged. Ultrasonic waves emitted from the active transmitter array propagate in a fan shape from the 1D array[14]. Ultrasonic waves reflected from the target are observed through the active receiver array. This measurement event is performed for all combinations ($N^2$) of the transmitter arrays ($N$ rows) and the receiver arrays ($N$ columns). This measurement principle allows the measurement of $N^2$ data points with $2N$ circuits. This schematic diagram has 8 rows and 8 columns for clarity.

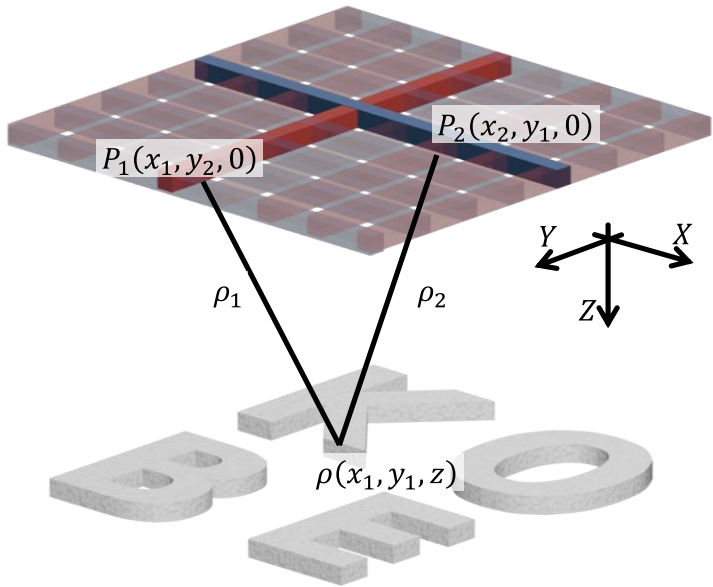

Figure 2 Mathematical concept of the measurement system used in this study.
On considering the wave emission and incidence at each point on the orthogonal linear array, we derive the imaging function for the entire array. Waves emitted from the point $P_1$ on the transmitter array reach an arbitrary point $\rho$ on the target through an arbitrary path $\rho_1$. Waves reflected from the point $\rho$ are detected at the point $P_2$ on the receiver array through an arbitrary path $\rho_2$. Based on the scattering field theory in 11-13, we derived the partial differential equation (1).

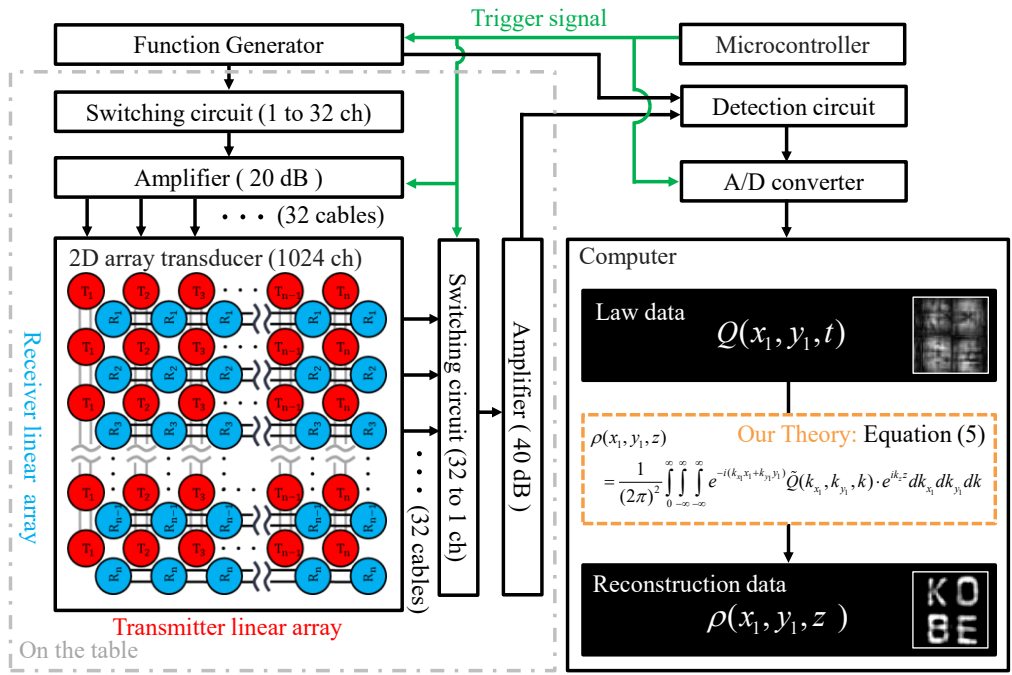

Figure 3 Block diagram of the image reconstruction process in the ultrasonic imaging system developed in this study.

The orthogonal linear array transducer uses an aerial ultrasonic sensor (MA40S4S/R, center frequency 40 kHz)[15], which has characteristics useful for principle verification. It is an orthogonal linear array system in which the transmitter array (row direction) and receiver array (column direction) are each one-dimensionally arranged. The transmitter array is driven by a 500-mV$_{pp}$ square wave from a Tektronix AFG1022 function generator via an INA110KP-based amplifier circuit (20 dB gain), with all elements emitting ultrasonic signal in-phase. The receiver array detects the reflected wave. The signals are detected using an INA110KP-based amplifier circuit (40 dB gain).

The switching circuit uses an ADG732 multiplexer, and the transmitting/receiving channels that can be selected for each row and column are controlled by a 5-bit digital signal. This structure allows to perform the measurements of $N^2$ the combinations of transmitters and receivers using circuit connections for a total of channels. The measured data can be substituted into the imaging function in Equation (5) derived by the scattering field theory to visualize the reflection intensity distribution at any depth $z$ mm.

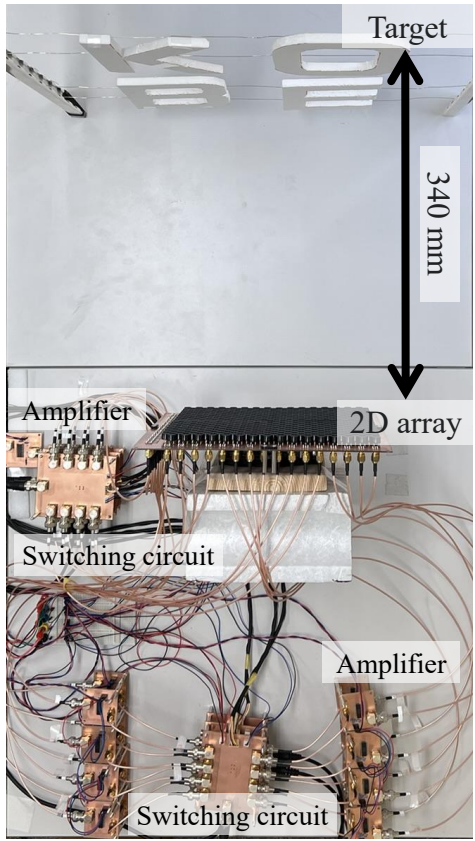

Figure 4 Experimental setup for imaging a target at a distance of 340 mm using the ultrasonic imaging system developed in this study.

The target was cut from a plate of foam material and placed in front of the orthogonal linear array transducer. In the experiment, equivalent measurements were achieved by scanning the 16 × 16 element array four times at different positions as an alternative to the 32 × 32 element array.

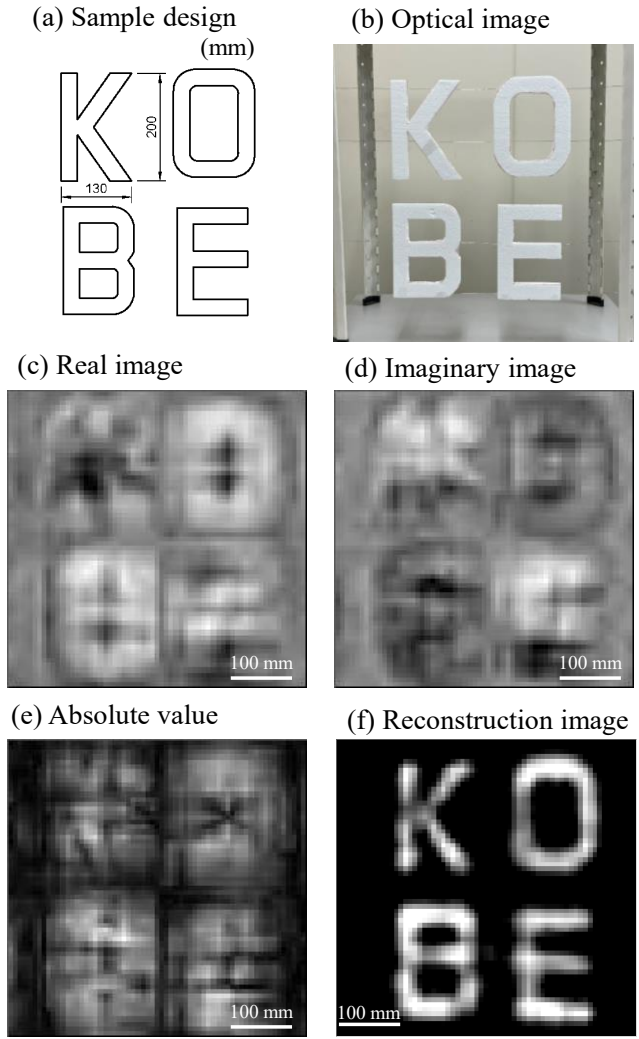

Figure 5 Experimental results using the ultrasonic imaging system.
(a) The size of the letters cut out of the foam material; (b) an optical photograph taken of the letters taken from the front; (c) and (d) the center frequency (40 kHz) components of the orthogonal biphasic data of the reflected wave obtained experimentally in the setup shown in Figure 4; (e) the absolute value of (c) and (d); and (d) the image obtained by calculating the reflection intensity distribution at z = 340 mm using the imaging function in Equation (5).